\documentclass[journal=jacsat,manuscript=article]{achemso}

\ifnum \value{page}=1
\usepackage{chemformula} 
\usepackage[T1]{fontenc} 

\author{No\"elie Duchamp, Chlo\'e Feschet, Maria M. Tarrag\'o and Philip E. Hoggan}
\affiliation{Institut Pascal, UMR 6602 CNRS, BP 80026, 63178 Aubiere Cedex, France.}
\email{philip.hoggan@uca.fr}

\title[An \textsf{achemso} demo]{Quantum Monte Carlo method modeling supported metal catalysis: Ni(111) converting adsorbed formyl 'en route' to hydrogen.}


\begin{document}

\begin{abstract}

Hydrogen production as a clean, sustainable replacement for fossil fuels is gathering pace. Doubling the capacity of Paris-CDG airport has been halted, even with the upcoming Olympic Games, until hydrogen-powered planes can be used.

It is thus timely to work on catalytic selective hydrogen production and optimise catalyst structure. Over 90 \% of all chemical manufacture uses a solid catalyst.
This work describes the dissociation of a C-H bond in formyl radicals, chemisorbed at Ni(111) that stabilises the ensuing Ni-H linkage. As part of this mechanistic step, gaseous hydrogen is given off.

Many chemical reactions involve bond-dissociation. This process is often the key to rate-limiting reaction steps at solid surfaces.
Since bond-breaking is poorly described by Hartree-Fock and DFT methods, our embedded active site approach is used. This work demonstrates Quantum Monte Carlo (QMC) methodology using a very simple monolayer Ni(111) surface model.

The rate-limiting reaction step of formyl decomposition to hydrogen and carbon is the initial C-H bond stretch. The full dissociation energy is offset by Ni-H bond formation at the surface. Reactive formyl (H-C=O) radicals also interact with a vicinal Ni. These adsorbed formyl radicals then produce carbon monoxide and hydrogen, with a H-atom dissociated from the formyl radical another desorbed at the Ni(111) face.



\end{abstract}

\section{Introduction}

\hskip5mm Hydrogen powered trains and cars are now in use and described by popular press \cite{SA}. Jet engines tested on hydrogen by Rolls Royce funded by airlines show the only remaining difficulty is storage requiring 4 times the space of kerosene, and condensation \cite{bbc}.

This work presents a simple model of Ni(111) using a periodic hexagonal monolayer, represented by the primitive cell. This is simply a lozenge of four Ni atoms, with 60° angles. This motif can be repeated periodically in two-dimensions. Above the surface, a formyl radical is adsorbed, bridging two Ni atoms and this system is converted to adsorbed carbon monoxide and H$_2$, the fuel product. The carbon monoxide is used in water gas shift hydrogen production, previously studied \cite{hog1}. The formyl species and optimum geometry for the Ni-monolayer were taken from that of the surface layer of Ni(111) obtained by Mavrikakis at the Perdew-Wang (PW91) Density Functional Theory (DFT) level \cite{Mav}. The reactant geometry was optimised with a second hydrogen atom that adopts an equilibrium position bridging the pair of Ni atoms in the lozenge opposite that bridged by the formyl radical and at a distance of 1.3 \AA \hskip2mm from it (see Figure 2, below).

Once a periodic DFT wave-function has been obtained using the ABINIT software \cite{abin}, it is written in a format that serves as the trial wave-function for Quantum Monte Carlo (QMC) calculations. This is expressed in Slater-Jastrow form and the parameters of a generic Jastrow factor are optimised variationally during Variational Monte Carlo calculations (VMC). This allows electron correlation to be taken into account. The ground-state of this reactant structure is obtained by long Diffusion Monte Carlo (DMC) runs.

Next, the properties, in particular the total energy are compared to those of the Transition state (TS). This structure is difficult to obtain accurately and represents the energy maximum along the reaction path and minimum in other directions. We have approached it with model geometries, which relax to reactants or products when the geometry is optimised. Identifying the reaction path between adsorbed formyl and adsorbed CO is not trivial. It is given by eigen-vector following, using QMC force-constants \cite{fark}. We then have a model geometry for the TS, obtained by imposing equal chances of reaching reactants and products when relaxed.

\section{Methods}

\subsection{Study of hydrogen abstraction from the formyl radical.}
First, consider the radical H-C=O, with one unpaired electron. This requires a spin-polarised wave-function (formally dominated by a doublet spin-state).  Extension of the C-H linkage, towards carbon monoxide and nascent hydrogen passes a minimum at a stretched bond-length, before increasing.
The calculation was carried out with at the Density Functional Theory (DFT) level, using the Perdew Burke Ernzerhof (PBE) functional and ABINIT software. The wave-function could be improved, with a larger plane-wave basis and configuration interaction (CI).
Next, we considered a system that is closed shell overall, with an extra hydrogen atom, fixed at a distance of at least 3.2 \AA \hskip2mm from the carbonyl carbon. When this position was opposite the dissociating C-H, as in the H$_2$C=O molecule (e.g., positive and negative x-values if the C=O is at x=0, {say along y}), stretching the C-H bond leads to a monotonic increase in energy. On the other hand, when both the H-atoms are on the same side of the C=O; x>0, say, then the energy penalty in stretching the C-H bond is rapidly offset by the formation of H$_2$.
Since placing a second H at a stationary position in this molecular system is artificial, we devised a simple catalyst to fix this atom. It is a monolayer representing Ni (111). The geometry of this solid surface is optimised, then kept constant. Four Ni-atoms define the primitive hexagonal cell.

\begin{figure}
\includegraphics[scale=0.4]{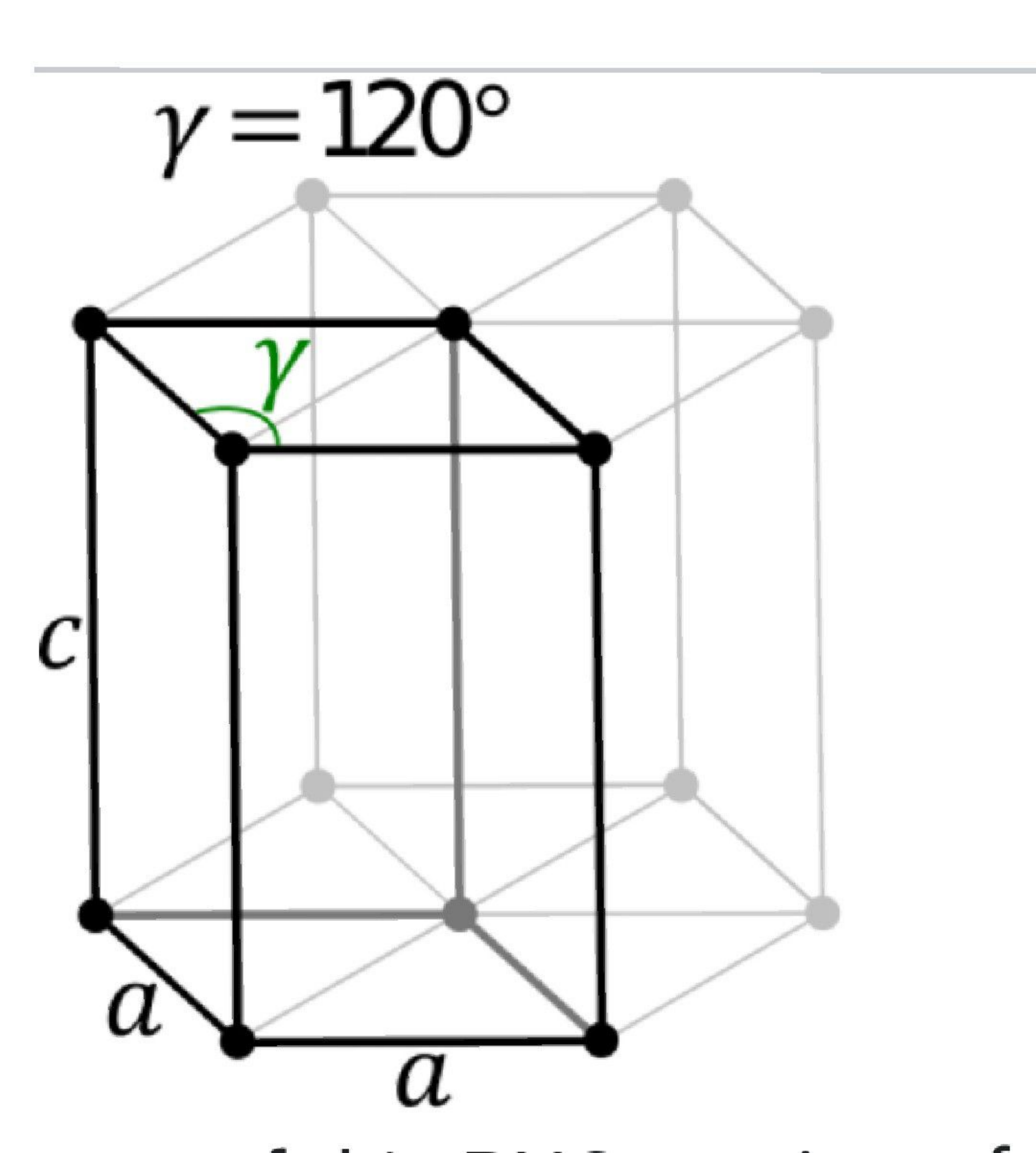}
\caption{Primitive unit cell for Ni, exposing (111) surfaces top and bottom. The lattice parameter a, is 6.68\AA \hskip2mm and our model represents this surface.}
\end{figure}

\vskip50mm
The first Ni-atom is placed at the origin with a neighbouring (vicinal) Ni-atom 4.6 bohr (2.142 Å) along the x-axis. At y = 4 bohr and indented 2.3 Bohr  the apex Ni of a Ni-atom equilateral triangle. Bridging these two Ni-atoms (origin, apex) is the adsorbed formyl radical.
This vicinal Ni atom pair is bridged by the second hydrogen, whilst the formyl group has a stretching C-H bond. Interestingly, such a structure offsets the energy of hydrogen abstraction to the extent that the bound structure of minimum energy has a considerably stretched C-H at 2.7 bohr (1.43 Å), where the dissociating hydrogen bridges two Ni-atoms. This is the case when a stretch along the x-axis is imposed, but is even more stabilised when stretching along y. Apparently, the reason for this is that the dissociating H-atom bridges two Ni-atoms, the second of which does NOT have a co-ordinated hydrogen already. Nevertheless, it is subsequently to be determined how the H$_2$ molecule will form (although H-atoms are very mobile), whereas the stretch along the x-axis gives a maximum around 3.2 bohr (1.69 Å), when the H$_2$ molecule can form, with the H-atom adsorbed at the vicinal Ni-atom.  The H-H distance decreases to 1.37 bohr (0.723 Å), close to the equilibrium H$_2$ bond length, although the atoms remain strongly bound to the Nickel (111) face.

The apparent barrier for this step is around 53 kJ/mol. The desorbed H$_2$ molecule is over 100 kJ/mol stabler.
This requires several comments on the model catalyst. It is a 2-D periodic system, only one Ni-atom thick, repeating the 4-atom motif that defines the (111) surface.
First, the potential energy surface around the adsorbed formyl equilibrium geometry is very flat. Whatever deformations are tested, the molecular system is stabilised by interactions with the Ni-monolayer.  This monolayer is more reactive than a catalyst surface, even a supported monolayer. The reason is that the under-side has no co-ordinated atoms, unlike bulk Nickel or supported Nickel. Reference work, in the case of platinum, shows that a supported Pt (111) monolayer is more reactive than bulk platinum, limited by Pt (111). The reaction studied was different but related: the attack of pre-adsorbed carbon-monoxide by water and we bench-marked the bulk Pt (111) at 71 kJ/mol1 and obtained a reliable estimate for the monolayer supported by Al (111) at 64.5 kJ/mol.
\vskip4mm
The author and his collaborators have consistently used Quantum Monte Carlo methods at chemical accuracy (1 kcal/mol=4.2 kJ/mol) for activation barriers of reactions at metal catalyst surfaces, since publishing test benchmarks on hydrogen dissociation on Cu(111) in 2015. Progress has been made in controlling error sources with the Casino QMC code for periodic systems by K. Doblhoff-Dier and P. E. Hoggan \cite{kdd1,kdd2}.
We previously worked on water-gas shift hydrogen production \cite{hog1}.

These studies describe water attack on pre-adsorbed CO at Pt(111). Our recent publication gives the associated rate-limiting barrier to within 1kJ/mol \cite{hjcp,shar}. This benchmark uses a Multi-reference (MRCI) trial wave-function, suited to bond breaking and formation and an embedded active site approach to the catalyst, allowing high-level input to be used for a small molecular active site, before the periodic QMC. PRACE 50Mh cpu was used \cite{prom}.

Other applications of QMC to catalysis and weak interactions of accuracy close to 1 kJ/mol are rare and often used to calibrate DFT functionals for cheaper investigations \cite{kdd3, oakr}.

Some reliable DFT results on catalytic hydrogen production are also noteworthy \cite{Camp,Faj2020}.

It clearly would be of interest to obtain valid information on the catalytic process with simpler wave-functions. Having tested the use of single-determinant (ground state) wave-functions for CO/Pt(111) it is used here, for a Ni(111) model. Limiting systematic error due to imperfect nodes from simpler trial wave-functions to 2.5 kJ/mol was studied in  \cite{hoggadv}. Barriers are over-estimated since transition states are strongly multi-reference.

Nowadays, stochastic methods are increasingly useful in
work on the challenges of real systems which require electron correlation accurately.
One such growth area for application is heterogeneous catalysis, in particular adsorbed reactions on metals.

In this work, we introduce the highly accurate Quantum Monte Carlo (QMC) approach for simple model active sites in a periodic solid.


This theoretical study of the initial C-H stretch step in formyl radical decomposition to produce hydrogen gas on  Ni(111) film catalyst follows several metal catalysis benchmarks by novel Quantum Monte Carlo methods, by the present authors \cite{hjcp}.

\subsection{The stages of our approach are as follows:}

1-Define a periodic 4 Ni-atom site, one receives H, the others form the equilateral triangle motif in the (111) face and the hollow site where H-C=O is adsorbed at the triangle centre of mass. Atom-list: Ni$_1$, Ni$_2$, Ni$_3$, Ni$_4$, C, O, H$_a$ and H$_b$. Placing the origin at Ni$_1$ and:
\vskip1mm
\begin{figure}
\includegraphics[scale=0.3]{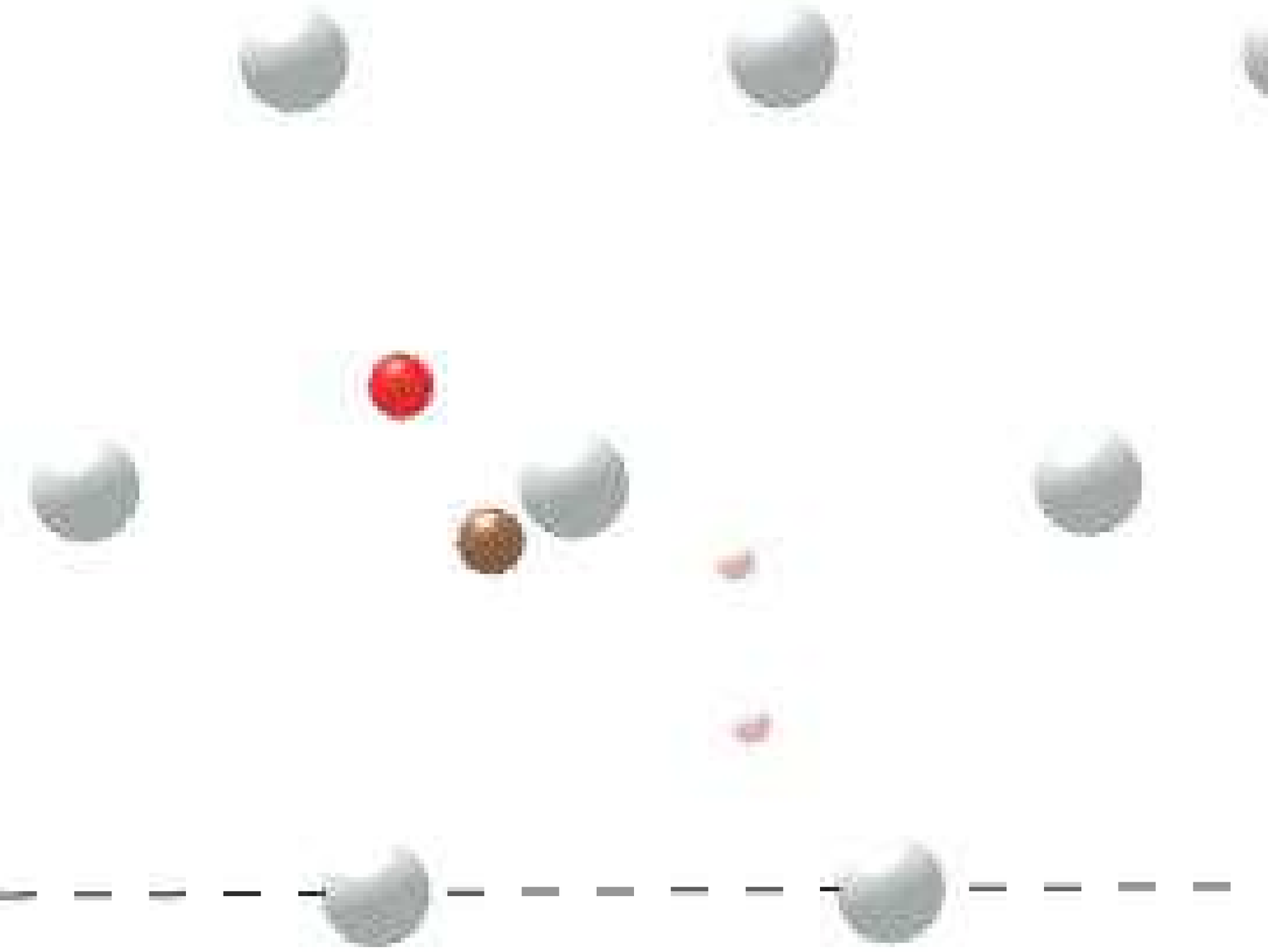}
\caption{Primitive unit cell for Ni, exposing (111) surfaces top and bottom. The lattice parameter a, is 6.68\AA \hskip2mm TS closeup.}
\end{figure}


  Ni$_2$,   4.6000 0.0000  0.0000

   Ni$_3$,   2.3000  3.9840  0.0000

   Ni$_4$,   -2.3000  3.9840  0.0000

   C,                   2.3000  0.6200  2.8220

   O,                   2.3000 2.0000  4.7704

   H (formyl),                    2.3000 -1.3700  4.0000

   Hb,                    2.300 -2.5500  1.9410

This Ni$_4$ defines a periodic (111-monolayer) slab. The Ni-slab wave-function is expanded in plane waves, with periodic boundary conditions in k-space.



3-The Slater determinants obtained from this wave-function are expressed in Slater-Jastrow form (as products with a generic Jastrow function).

4-This trial wave function with arbitrary parameters (polynomial coefficients of inter-particle distances in the Jastrow factor and Slater determinant weights) are optimised using Variational Monte Carlo (VMC).





5-This wave-function is then used to generate a population of 'walkers' (configurations) that are propagated in imaginary time during the second, diffusion step (DMC). DMC is carried out in the fixed-node approximation, keeping nodes from the input trial wave-function. An updated overview of the VMC and DMC methods is given in \cite{cas2020}. These trial wave-functions can be optimised with a complex Jastrow factor, \cite{garnet}  because they potentially provide the input with exact nodes. This improves the single Slater determinant \cite{geo} describing a ground-state from the DFT orbitals for heterogeneous systems. DFT nodes may well be poor.


CASINO software \cite{cas2020} is used but cannot be treated as a 'black box'. Many choices need user guidance.

A key step of an industrially important reaction is described in the present work: rate-limiting hydrogen abstraction from formyl.
This occurs on a Ni(111) film. It converts adsorbed (bridging) H-C=O, formyl to carbon monoxide (CO) and hydrogen. The toxic CO may then be used to produce more hydrogen in catalysed water-gas shift processes.

\section{Theoretical Method:}

\hskip5mm Inclusion of semi-core electrons in the valence is certainly necessary when dissociating molecules involving the transition metal atoms. It also improves test results for platinum. The copper 3d$^{10}$ shell is dense in the core-region but platinum has its 5d$^9$ 6s$^1$ ground-state shell on average further from the core. Pt is thus less prone to difficulties defining pseudo-potentials. Ni has a 4s$^2$ 3d$^8$ ground-state and requires semi-core electrons to reduce the non-locality pitfalls observed for copper\cite{kdd1}. They is limited here by using the Casula algorithm \cite{casula}.
Comparing asymptote and QMC optimised TS geometries for reaction barriers also involves the same atoms, limiting PP errors. Co-ordination of the TS to the metal surface occurs but the distance between atoms is about 40 \% more than the equilibrium bond--length and so the role of semi-core electrons is minor.
A Z=18 core for Ni is validated \cite{fpseu}.

\subsection{Variation Monte Carlo:}

\hskip4mm A preliminary Variational Monte Carlo (VMC) calculation is carried out in order to generate several thousand configurations (instantaneous points in electron co-ordinates). VMC is driven by minimisation. The so called {\it local energy} is minimised :

$${H \psi } \over \psi$$

The kinetic energy terms involved are smoother and have lower variance in exponentially decaying bases, as shown in our work on wave-function quality \cite{wfq,tlse}. A Jastrow factor including electron pair, electron-nucleus and three-body (two-electron and nucleus) is defined. This Jastrow factor \cite{asy,con} is carefully optimised (essential work, taking up to 5 \% of the total time). This factor uses a polynomial expansion in the variables of explicit correlation.


The product of a Jastrow factor with a Slater determinant gives the trial wave-function.
\hskip4mm The method of choice must scale well with system size, be able to efficiently use modern supercomputer facilities that are massively parallel and, above all, produce quantitative, accurate physical properties that are difficult to obtain otherwise. QMC benchmarks of activation barriers for reactions adsorbed on solid catalyst surfaces fit this description well.
Developing and applying QMC calculations on this scale require access to a supercomputer.

The Quantum Monte Carlo approach uses statistical physics over a large population, comprising sets of instantaneous particle positions in co-ordinate space. They are often called 'walkers' (c.f. the one-dimensional random-walk. Random numbers actually serve to initialise the 'walkers' from the trial wave-function giving initial electron density/population).

This method takes a Slater-Jastrow trial wave-function, which is optimised with respect to its polynomial coefficients during the variational (VMC) stage.
The arguments are the electron-electron distance for populations of instantaneous configuration in the first term. The second is the corresponding polynomial for electron-nuclear distance and the third for electron pair and nucleus (three body) nested sums over e-e and e-n distances. This last term is essential and needs to be treated with care. There is some 'double-counting' with the previous terms and it can lead to a very large parameter set with individual atoms treated (as in this work) and expansion to order greater than 3 (4 in this work). Comparisons with all four Ni-atoms grouped and also with third order expansion show better Variational convergence but large population explosion and deep sub-variational DMC in some cases. This has to be avoided by modifying its structure but is a common effect of non-local pseudo-potentials for 3d electrons, particularly (Ni is a 3d transition metal, see \cite{kdd1}).

These systems involve stretched bonds requiring almost all the electronic correlation. QMC has proven to be a reliable method for describing stretched bonds \cite{boufh}.

Generic Jastrow factors were used for this molecule-solid system (with 8 terms 1/atom). Note that cutoff functions are switched on to avoid the tree-body term double-counting some contributions already included in electron-electron and electron-nucleus terms.

We define it with just the half-axis z<0 'outside' the catalyst that includes the adsorbed molecules.

Note that z>0 is perpendicular to the thin film Ni(111)surface.
The x and y dimensions should be much longer than the maximum bond-length to limit finite-size effects.
The surfaces in this model are planar, whereas it is known that metal close-packed faces re-arrange forming high Miller index ridges or meanders. \cite{pimp} Our planar-surface model uses the transition state (TS) and
a reference state with the same atoms as the TS, before the reaction.

\subsection{Diffusion Monte Carlo:}

\hskip4mm In the DMC method the ground-state component of the trial wave function is projected out by solving the Schr{\"o}dinger equation (SWE) in imaginary time. DMC is carried out in the fixed-phase approximation, which uses the nodes from the input trial wave-function. An updated overview of the VMC and DMC methods is given in \cite{cas2020}. These trial wave-functions can be optimised with a complex Jastrow factor, \cite{garnet}  because they potentially provide the input with exact nodes. This improves the single Slater determinant describing a ground-state from the DFT orbitals for heterogeneous systems. DFT nodes may well be poor.
The imaginary-time SWE is transformed into a Diffusion equation in the 3N-dimensional space of electron coordinates by replacing $t$ by the pure-imaginary time $it$.

\section{Results}

\hskip5mm For the formyl moiety (H-C=O) adsorbed on a Ni-film the C-H dissociation barrier is evaluated (towards CO release): A local minimum (reaction intermediate) geometry was tested to resolve the energy difference compared to starting (reactant) geometry comprising the same atoms, from DFT-PW91 work by Mavrikakis \cite{Mav}.

We obtain: DFT-PBE difference of 48.9kJ/mol VMC 71kJ/mol, and DMC 53.45kJ/mol with a standard error (se) below 2kJ/mol.

After further investigations, which involve following the reaction path, a likely TS geometry candidate appears 97.5 kJ/mol above a slightly optimised starting geometry using DFT-PBE. This structure desorbs a formyl radical (which readily saturates with hydrogen), when optimised, or adsorbed carbon-monoxide and H$_2$ (products). this clearly is typically TS behaviour. The geometry has been confirmed determining force-constants to identify the reaction-pathway, using algorithms developed in \cite{fark} which is now available within QMC.

The VMC barrier here is 105.2 kJ/mol and DMC 90.4 kJ/mol, with se 2.15kJ/mol. 

Here, 30000 DMC stats for each geometry converge it, to an error in se ratio R <1/10.

The model used has no co-ordination below the four Ni-atoms defining the surface layer. Adsorbed molecules provide co-ordination above it, however, the missing interactions with the solid which is located below in the actual system is absent. This defect is compensated to a greater extent in the TS compared to the equilibrium geometry of pre-adsorbed starting materials (adsorbed formyl radical). Hence, it is not surprising to see that the DMC lowers the total energy of the TS system more than that of the reactants. This leads us to a DMC activation barrier that is lower than the DFT value (90.4 c.f. 97.5 kJ/mol).
The surface hence appears more reactive that the actual metal catalyst, with explicit bulk Ni and film.

It has also become clear that these DMC runs include a systematic error, also related to the 4-atom Nickel catalyst model. It stems from the Ni-pseudo-potential that has significant non-locality \cite{kdd1}. Nevertheless, the population of walkers only fluctuates by about 10 \% and the R threshold for converged runs at 1/10 is reached by 25-30 k data-points with no time-step bias and blocks of 256 points in averaging.

\section{Conclusions and perspectives}
\hskip5mm This work with a simple Ni$_4$ unit-mesh describing a monolayer surface, adsorbing H-C=O, but with no bulk below it. A single-determinant ground-state initial wave-function is used, comprising Kohn-Sham spin-orbitals gives credible results (the DMC barrier obtained in this way is 90.4 c.f. 92-97 kJ/mol from reference converged periodic grid DFT). The DFT total energies are used as control-variate in the QMC twist-averaging process \cite{cas2020}. Nevertheless, it is clearly desirable to use CI wave-function input, that will be the subject of future work.

The Ni (3d$^8$ 4s$^2$) pseudo-potential has some non-locality \cite{kdd1}. No redox phenomena are observed, therefore, this partly occupied 3d-shell is necessarily valence.

\vskip8mm
{\bf Acknowledgements.}

We are grateful the GENCI special allocation on Irene (nov 2022 to nov 2023 CEA, Bruy\`eres-le-Ch\^atel, F) and University of Clermont Mesocentre (data preparation, tests).

PEH thanks Pablo Lopez-Rios for helpful discussions.



\end{document}